\documentclass{kluwer}    

\usepackage[dvips]{graphicx}

\newcommand{\microns}{\ensuremath{\mbox{ $\mu$m}}}

\begin{document}                                                                \begin{article}
\begin{opening}         
\title{Probing the close environment of young stellar objects with interferometry} 
\author{Fabien \surname{Malbet}\email{Fabien.Malbet@obs.ujf-grenoble.fr}}
\institute{Laboratoire d'Astrophysique, Observatoire de Grenoble, UMR
  5571 CNRS/UJF, Grenoble, France}  
\begin{ao}\\
  Laboratoire d'Astrophysique, \\
  BP~53, \\
  F-38041 Grenoble cedex 9,\\ 
  France
\end{ao}

\begin{abstract}
  The study of Young Stellar Objects (YSOs) is one of the most
  exciting topics that can be undertaken by long baseline optical
  interferometry. The magnitudes of these objects are at the edge of
  capabilities of current optical interferometers, limiting the studies
  to a few dozen, but are well within the capability of coming large
  aperture interferometers like the VLT Interferometer, the Keck
  Interferometer, the Large Binocular Telescope or 'OHANA. The milli-arcsecond
  spatial resolution reached by interferometry probes the
  very close environment of young stars, down to a tenth of an
  astronomical unit. In this paper, I review the different aspects of star
  formation that can be tackled by interferometry: circumstellar
  disks, multiplicity, jets. I present recent observations
  performed with operational infrared interferometers, IOTA, PTI and ISI, and
  I show why in the next future one will extend these studies
  with large aperture interferometers.
\end{abstract}
\end{opening}           

\section{Introduction}

When trying to understand the origin of our planetary system, one has
basically two approaches: (i) looking for other existing
planetary systems in the universe to characterize them or (ii)
investigate how stellar systems have been forming. By studying young
stellar objects (hereafter YSOs) in our Universe, i.e.  stars in their
early stages of evolution, one focuses our attention to the second
approach.

So far, these objects have been extensively observed by
spectrophotometry with seeing-limited resolution that corresponds at
best to a hundred of astronomical units for typical star formation
regions, thus many questions are still \emph{unresolved}\footnote{Pun
  intended as pointed out by JDM} because our incapability to
disentangle the various phenomena at smaller scales.  The advent of
millimeter-wave interferometry, adaptive optics imaging and space
telescope observations in the last decade allowed us to probe these
systems closer to the star (down to 10~AU) but not close enough to
probe inner solar system scales.  Therefore long baseline optical
interferometry with milli-arcsecond resolution corresponding to a
tenth of a Earth orbit radius is an essential tool to unravel the
nature of the physical phenomena occuring in the early stage of star
and planet formation.

In this paper, Sect.~\ref{sect:har} briefly reviews the current
knowledge about YSOs, then Sect.~\ref{sect:small} describes the first
results obtained with presently operational interferometers and
finally Sect.~\ref{sect:large} introduces some expected science
objectives that will result from interferometric facilities with large
apertures that are being built, such as the VLT Interferometer, the
Keck Interferometer, the Large Binocular Telescope or 'OHANA.

\section{Stellar formation at high spatial resolution}
\label{sect:har}

The star formation process of solar-type stars lasts about 10~Myr. The
scenario of star formation has been categorized in classes of objects
corresponding to different stages of evolution \cite{And94} from class
0 to class III.  During the first 10,000~yr, the about 10~pc-wide
clouds at the origin of stellar systems collapse as a result of
reaching their critical mass.  Since the clouds are initially
rotating, they fragment in flattened rotating structures called
pre-stellar class 0 cores and observed at the sub-millimeter
wavelengths \cite{Scuba} since due to the large opacity the energy
released during the collapse cannot escape in the visible domain.
After about 100,000~yr, the collapsing envelopes around class~I
protostars dissipate and reveal the inner part of class~II systems
composed of a protostar usually surrounded by an active accretion disk
and a strong bipolar jet. The systems are now shining at shorter
wavelengths in the infrared domain, since the main sources of
radiation are the disk and the young stars still contracting.  At that
stage, the extent of the protoplanetary system reaches up to 1500~AU.
Later on, at about 1~Myr, the gas reservoir has dissipated and the
accretion stops. The class~III stars, with some planetesimals and
forming planets orbiting around it, are now shining mainly in the
visible, a tenuous disk remains when the star arrives on the main
sequence, resulting from the collisions of the planetesimals. In this
paper we focus our attention on class~II systems, 
targets of interests for optical interferometry.

The first evidence for the presence of circumstellar disks around
young stars came from the advent of infrared detectors and subsequent
surveys \cite{Men66,Men68} which culminated with the IRAS survey in
the early 1980's, as well as from the first UV space observations
(IUE). Indeed young stars exhibit an important UV and IR excess that
can even exceed the stellar luminosity.  Using disk models originally
developed by \inlinecite{SS73} and \inlinecite{LBP74}, several teams
(e.g.\ \opencite{BBB88}) showed that these excesses came respectively
from the radiation of the circumstellar matter orbiting at lower
effective temperature in the disk and from the interaction between the
disk and the slowly rotating star. These models work well for many
young systems, but for some, the spectral energy distributions (SEDs)
display even larger excesses, explained by passive flaring disk models
\cite{ASL88}.

Millimeter-wave interferometry succeeded in resolving some of the systems
with sub-arcsecond resolution (see for example UY Aur rotating disk
revealed by \opencite{Dut98}). At about the same time,
adaptive optics on 4-m class telescopes and the Hubble Space Telescope
(HST) directly brought to view such disks at visible and near-infrared
wavelengths (see e.g.\ HK~Tau B by \opencite{Sta98} and
HV~Tau by \opencite{MB00}). In fact most of these disks
are observed edge-on, because the contrast between the star and the
disk is decreased by the absorption of the stellar light by the
central part of the disk. The disk is seen in stellar light,
scattered in the upper part of the disk atmosphere. The extent of
these systems is about 100-200 AU with typical spatial resolution of
10 to 20 AU.


The Herbig-Haro 30 object observed by the HST in forbidden lines and
in the continuum \cite{Bur96} brought a lot of information. The
relevant region of investigation lies within the 10 AU from the center
where the accretion disk interacts with the star and the base of the
bipolar jet is forming. One does not know precisely how these
different phenomena work and interact. Is the accreting matter falling
onto the star through the equatorial plane or along magnetosphere
field lines?  Is the jet engine the rotating star or the upper part of
the disk atmosphere?  Where do planets form: in the outer part of the
disk or at the locations known for the solar system? What is the
physical phenomenon at the origin of the viscous dissipation: pure
hydrodynamical turbulence or manifestations of the magnetic field? What
is the disk chemistry? What is the spatial distribution of the matter?
All these questions can be investigated by visible and infrared
long-baseline interferometry.

The current capabilities of optical\footnote{When using the term
  \emph{optical}, I refer to both the \emph{visible} and
  \emph{infrared} wavelength domains.} long baseline interferometers
(with [0.5--10\microns] spectral coverage, baselines up to several hundreds
of meters) gives us the characteristic range of spatial scales:
between 0.1 and 10 AU, and a temperature range between few 100 K to
5000 K. With these numbers in mind, one can list in more detail what
can be investigated by optical interferometry:

\begin{list}{--}{\setlength{\itemsep}{0pt}}
\item Star surfaces: apparent diameters of young stars are typically
  0.1 mas, i.e.\ they will be barely resolved. In addition the
  circumstellar matter around them might pollute diameter
  measurements. However weak-line T Tauri stars, which are class~III
  stars, are believed to have lost their accretion disk and therefore
  some determination of stellar parameters like the radius, effective
  temperature or even limb-darkening might be possible.
\item Multiplicity: about 80\% of young stellar systems are expected
  to be multiple \cite{RZ93}. The binary frequency has so far been
  investigated in the domain of spectroscopic binaries; and
  visual and adaptive optics binaries, and interferometric
  observations will fill the gap between the two. Using
  interferometric data to get the orbit of known spectroscopic
  binaries will allow to derive their mass accurately.
\item Circumstellar disks: optical interferometry allows to study the
  thermal emission of the disks rather than the stellar light
  scattered by the disk. Therefore one has access to quantities like
  the temperature and eventually their surface density distributions.
\item Accretion: in the current paradigm, accretion occurs in a
  Keplerian disk. With some spectral resolution, one will
  verify this motion and even measure departure from the Kepler law
  (see Sect.~\ref{sect:kepler}).  The terminal accretion is not yet
  understood and neither is the link with the star magnetosphere. With
  spectral and spatial resolution, one can solve the
  kinematics of the inner region.
\item Ejection: one does not know yet if the origin of the jet comes from
  the stellar dipole/quadrupole or from the disk. Interferometry will
  allow to probe the base of the jets and determine their opening
  angle, which is a critical parameter (see Sect.~\ref{sect:jets}). One
  can also investigate the link between the jets at small scales
  and larger outflows.
\item Formation of planets: currently giant planets are believed to be
  born in the outer region of active accretion disk and to then
  migrate inward closer to the star. Such phenomenon perturbs the disk
  and may leave detectable signatures like the presence of gaps or
  gravitational waves in the disk.
\item Star formation scenarios: YSOs correspond to a collection of
  different objects at different times of evolution ranging from
  classical T Tauri stars to weak-line T Tauri stars , and also
  include FU Orionis objects which are believed to be T Tauri stars
  for which the disk undergoes an accretion outburst, and,
  intermediate mass counterparts called Herbig Ae/Be stars. Observing
  these different classes of objects will give us clues on the
  evolution of young stars.
\end{list}
Most of these phenomena takes place in a sub-AU region where the
dynamic time scale is less than a year for solar mass stars and even
shorter for more massive stars. Therefore one should also consider
that the investigated physics will no longer be stationary but for a
large part dynamical.

Finally, I would emphasize that observing YSOs will bring us a wealth of
information also  useful for other types of astrophysical
objects such as active galactic nuclei, cataclysmic variables,
radio-jets,... since it will shed lights on universal physical
processes like the nature of viscosity, the role of magnetic field,
how dust grains form and grow, and the link between accretion and
ejection processes.

\section{First results from operational small aperture optical
  interferometers}
\label{sect:small}

In this section, I summarize the results obtained so far in optical
interferometry on the brightest objects and I show that interferometry
has already modified our understanding of disks.

These results come from three infrared interferometers: the
\emph{Palomar Testbed Interferometer (PTI)} located on Palomar
Mountain (California, USA) and operating two pairs of siderostats in the $H$
and $K$ bands; the \emph{Infrared and Optical Telescope Array (IOTA)}
located on Mount Hopkins (Arizona, USA) operating in the $H$ and $K$
bands with two relocatable siderostats; and the \emph{Infrared Spatial
  Interferometer (ISI)} located on Mount Wilson (California, USA) and
operating at $10\microns$ with a heterodyne detection.

The list of refereed papers that describe all YSO observations
carried out todate is: FU Ori at PTI \cite{Mal98}; AB Aur  at IOTA and
PTI \cite{Mil99}; T Tau, MWC 147, SU Aur and AB Aur at PTI
\cite{Ake00}; about 15 Herbig Ae/Be stars  at IOTA \cite{Mil01}; and
the massive star LkH$\alpha$~101 at ISI \cite{Tut02}.

Most of the objects observed so far are either intermediate mass
objects or low-mass stars with an active accretion disk, because their
magnitudes are close to the sensitivity limit of the instruments. The
measurements are limited to visibility amplitudes and therefore are
mainly focused on the disk physics.

\subsection{Constraint on T Tauri disks}

One interesting example is the case of FU Ori. This system is known to
be composed of a T Tauri star with a disk undergoing an accretion
outburst \cite{HK96}. Therefore this system is brighter by more than
2-4 magnitudes than other classical T Tauri systems and the disk
emission dominates the spectral energy distribution. This object is
therefore an ideal target to study disk accretion, since at
near-infrared wavelengths the observations are not contaminated by other
physical processes.

\begin{figure}[t]
  \centering
  \includegraphics[width=0.7\hsize]{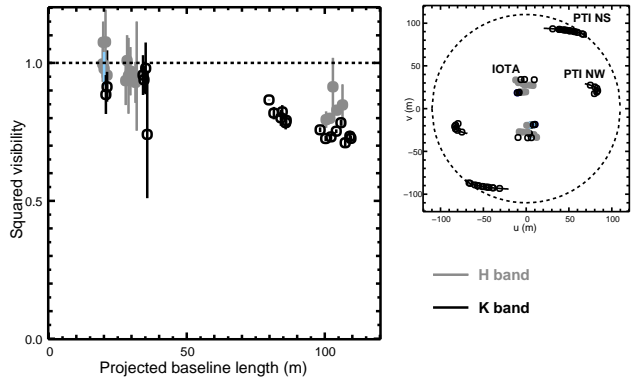}
  \caption[]{Interferometric observations \cite{Mal98,MB01} of FU Ori
    collected at IOTA and PTI between 1997 and 2000. Gray circles
    correspond to $H$ band, and black circles to $K$ band. Left panel:
    squared visibilities as a function of the projected baseline.
    Right panel: $(u,v)$ coverage.}
  \label{fig:fuori}
\end{figure}
Figure \ref{fig:fuori} shows a typical example of data collected with
IR interferometers. The visibility reduction of 15-30\% at long baselines
corresponds to the resolution of a structure of size of the
order of 1~mas, corresponding to 0.5~AU at the distance of the
Orion cloud. Moreover, the visibility drop in $H$ is
smaller than in $K$, meaning that the size of this structure
decreases with the wavelength faster than the resolution $\lambda/B$.

The data is consistent with the standard disk model used to fit the
SED. In fact the data points are lying exactly where they were
expected to be from disk model predictions \cite{MB95}.

However the data collected in the following years in the $H$ and $K$
bands show square visibilities which differ more than expected by the
standard accretion disk model.  A way to investigate this departure
from the standard model is to derive a direct measure of the
temperature law. \inlinecite{ASL88} have introduced an \emph{ad-hoc}
prescription to describe the variation of the effective temperature
through the disk:
\begin{equation}
  T = T_0 \, \left( \frac{r}{r_0} \right)^{-q} 
  \quad \mbox{ with } \quad 0.5 \leq q \leq 0.75
\end{equation}
A value of $q=3/4$ corresponds to the standard geometrically thin
disks and $q=0.5$ describes some type of flaring passive disks.
\inlinecite{BS90} surveyed a large number of sources at millimeter
wavelength and could derive values of $q$ which indeed range between
0.5 and 0.75.

In the case of FU Ori, \inlinecite{Mal02} obtained $q \simeq
0.6\pm0.2$, compatible with the SED derived value of $q \simeq 0.75$.
However, in the future more precise observations might no longer be
interpreted with this simple disk model. The reason for such behavior
can be explained by the fact that interferometry probes the inner
region of the disk whereas the SED is more sensitive to the outer
regions. One speculates that the change of temperature law is due to
a change of the accretion rate in the inner part of the disk due
to the presence of a jet that removes material from the disk.

The case of FU Ori is simple since the dominant physical process is
the accretion. In low-mass stars YSOs, some other effects can play a
substantial role, like the heating of the disk photosphere by the
central star \cite{Cal91,MB91,CG97}. This creates an upper layer
which can have higher temperature than the layer located at lower
altitudes, impacting both the SED and the visibilities. 

To describe this stellar reprocessing in the disk, there exists no
simple models. \inlinecite{CG97} presented a two-layer model that is
successful in interpreting SEDs. More recently \inlinecite{LMM03}
generalized the \citeauthor{CG97} model to include the effect of
viscous dissipation.  They have applied their model to interferometric
observations made by \inlinecite{Ake00} at PTI and succeeded in
fitting reasonably well both SED and visibilities. It therefore shows
that interferometric data can provide valuable information on the physics
of the disk.

\subsection{Disks around Herbig Ae/Be stars are not as expected}

Herbig Ae/Be stars (HAEBEs) are young stars of intermediate mass (a
few solar masses). They exhibit an infrared excess like the lower mass
T Tauri stars, but there has been a controversy on the origin of this
excess \cite{Hil92,BC94}. Since they are brighter than T Tauri stars,
they have been excellent candidates to be observed by infrared
interferometers. 

\inlinecite{Mil99} observed the first Herbig Ae star
AB Aurigae in 1997-1998 and discovered that the standard disk model was
not compatible with the data. In brief, the object was over-resolved
compared to the prediction of the model. This result was confirmed by
observations of about 15 targets of the same type by \inlinecite{Mil01}. 

The best interpretation of the data is to consider that the star is
surrounded by a ring of material located at sub-AU distances, namely
0.3~AU for AB Aur. \inlinecite{Tut01} obtained similar results on the
more massive star LkH$\alpha$~101, and proposed that this ring is the
result of the sublimation of the dust orbiting the star, such that
the ring radius gives an estimate of the distance where dust
sublimation occurs.


\inlinecite{MM02} compiled measurements made with the various
interferometers and put them in a diagram with the ring inner radius
as a function of the central star luminosity. The correlation seems to
follow the law that can be expected by a simple model of dust
sublimation in $r^{-2}$. In parallel, \inlinecite{Nat01} who worked on the
SEDs measured by ISO were not able to reproduce the gap occuring in
the near-infrared with the standard model. By truncating the inner
part of the disk and heating the inner part of the disk by the central
star, they were able to fit both the ISO data and the interferometric
observations. The ring proposed by Millan-Gabet and collaborators
would then be the inner part of a disk which is puffed up by the
stellar radiation and may also shadow part of the disk (see also the model proposed by \opencite{Dul01}).

It is less clear whether T Tauri stars follow the same behavior. As a
matter of fact, the regions for dust sublimation models and accretion
disk are very close for these objects. More detailed studies are therefore
required for these low-mass objects and should be possible with large
aperture interferometers.

\subsection{Conclusions and perspectives}

One starts constraining the physical conditions in disks with
different techniques (millimeter-wave interferometers, adaptive
optics, HST, optical interferometry), but optical interferometry is
the only current technique which constrains the inner 10~AU of the
disk. The first results on low-mass T Tauri stars are consistent with
standard accretion disk models, but more complex environments appear
to be required for the more massive HAEBE stars with passive disks.

One needs to improve the wavelength coverage since it is a good way to
constrain temperature laws. Ideally $J$, $H$, $K$, $L$ and $N$ data
sets and better $(u,v)$ coverage are desirable.

The coming facilities with large apertures like the VLTI, KI, LBT and
`OHANA will certainly boost these studies by bringing higher
sensitivity, better $(u,v)$ coverage, access to phases and eventually
to images. Finally, it is also essential to continue the theoretical
modelling effort, particularly relevant physics of the inner disks.

\section{Expected science from coming large aperture interferometers}
\label{sect:large}

Some very interesting science has been carried out with current
interferometers with small apertures. In this section, I present in
the YSO field the impact of anticipated improvements that will be
accessible with interferometers with large apertures like the VLTI,
the Keck interferometer, the LBT and 'OHANA.

\subsection{Improvements expected from large interferometers}

\begin{figure}
  \centering
  \includegraphics[width=0.8\hsize]{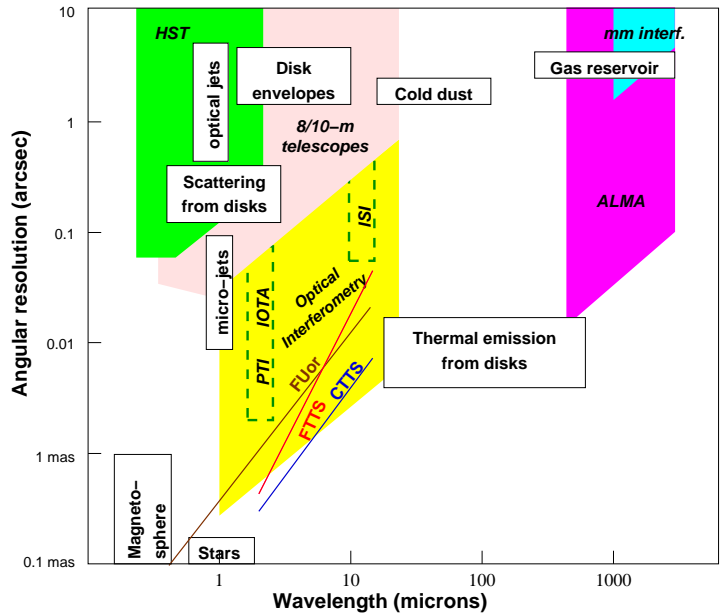}
  \caption[]{Resolution diagram showing angular resolution as a function of 
    wavelength. Competitive instruments from optical to millimeter
    wavelengths are displayed together with the main issues in the
    star formation field. Also present are the operational infrared
    interferometers that have already produced some astrophysical
    results: IOTA, PTI and ISI.}
  \label{fig:resolution}
\end{figure}
\begin{figure}
  \centering
  \includegraphics[width=0.48\hsize]{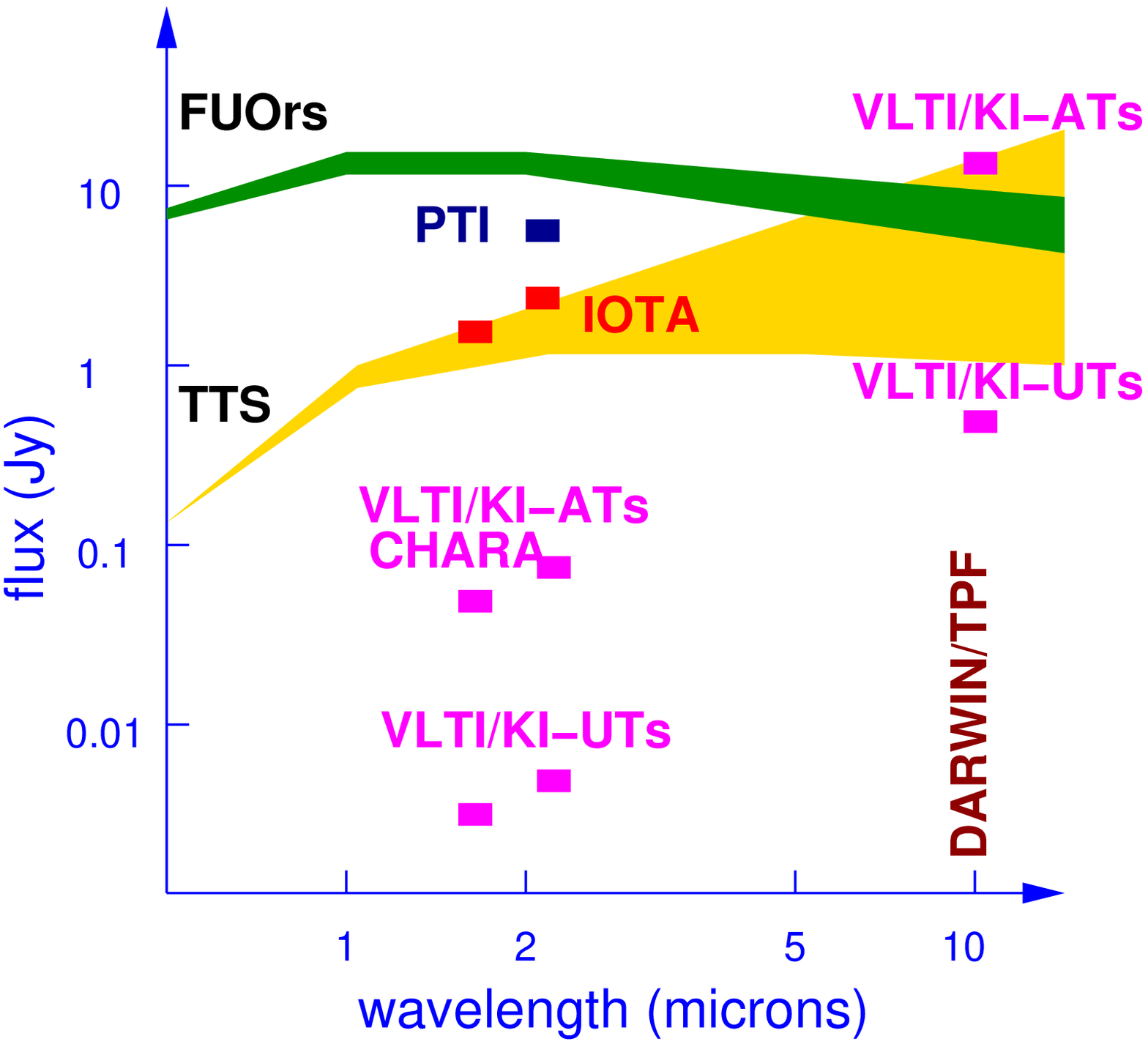}
  \hfill
  \includegraphics[width=0.4\hsize]{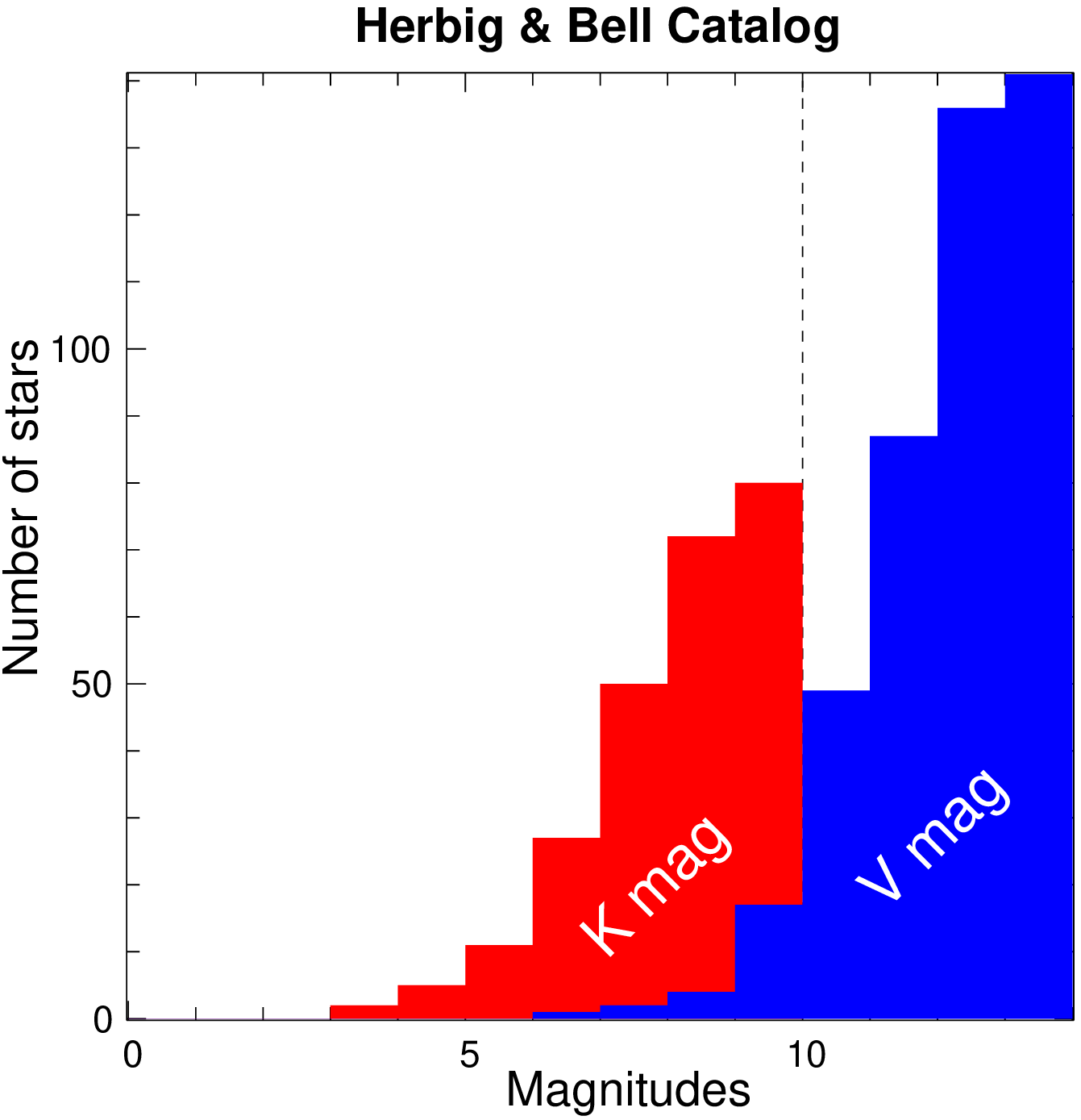}
  \caption[]{Left panel: Sensitivity diagram showing the flux sensitivity as a
    function of the wavelength in the optical range. Sensitivities
    from IOTA and PTI are represented as well as the typical flux for
    T Tauri stars (TTS, both classical ones and those with a flat IR
    spectrum) and FU Orionis stars (FUOrs). Right panel: $K$ and $V$
    histograms of objects listed in the  Herbig \& Bell
    catalog \cite{hbc}. A limiting magnitude of $K=10$ (dashed line) is
    a conservative value for an interferometer with large apertures.}
  \label{fig:sensitivity}
\end{figure}

The angular resolution and the spectral coverage offered by presently
operating interferometers match almost perfectly the needs in the star
formation domain (see Fig.~\ref{fig:resolution}). YSOs are usually at
the limit of detection of present small aperture telescopes with a
typical $K$ limiting magnitude of 6.  Only very few objects can be
observed and are mainly the more massive ones like Herbig Ae/Be stars (see
results presented in Sect.~\ref{sect:small}).

Usually in interferometry, the magnitudes of the objects are not the
unique detection criterion but the expected visibility is also
important: interference fringes must be detected and the amplitude of
these fringes are proportional to the correlated magnitude, i.e.  both
to the flux and to the visibility amplitude at the given spatial
frequency. However in young stars, since most structures are barely
resolved, one does not expect the correlated magnitude to be smaller
by 1 or 2 magnitudes than the object magnitude and the object
magnitude remains a good criterion. The left panel of
Fig.~\ref{fig:sensitivity} displays the spectral energy distribution
for different types of low-mass stars located respectively at 150~pc
and 450~pc: classical T Tauri stars, T Tauri stars with a flat
infrared excess and FU Orionis objects. HAEBE stars and more massive
stars usually lie far above these curves. On the same figure, I
indicate the sensitivity limits of PTI and IOTA demonstrating
that these interferometers are at the very edge of the domain.
Expected performances from interferometers using 2m (ATs) and 8-10m
(UTs) class telescopes like the VLTI, KI, LBT or `OHANA show that
those interferometers are sensitive enough to fully investigate these
objects at least in the near-infrared. In fact the right panel of
Fig.~\ref{fig:sensitivity}, which plots the $K$ and $V$ histograms of
the young stars listed in the Herbig \& Bell catalog \cite{hbc} shows
that with a $K$ limiting magnitude of 10, a conservative value,
several hundreds of objects are observable.  With such a high number
of potential targets, one can now consider carrying out surveys in
order to perform statistical studies where one would look for
correlations between visibilities and flux excesses, binary
frequency, or veiling.

Another limitation of presently operating telescopes is the number of
operational baselines. IOTA is limited to a maximum baseline length of
38m, PTI can only use long baselines of 80m and 100m (see for
example the $(u,v)$ coverage of FU Orionis shown in
Fig.~\ref{fig:fuori}).  This is not sufficient to obtain a complete
characterization of the resolved material. In the near future, the
VLTI will be able to provide a wide coverage of a 200-m large
synthetic aperture thanks to 4 relocatable auxiliary telescopes. CHARA
with a maximum baseline of 350m even if with only 1-m apertures will
provide more angular resolution. 'OHANA with a limited number of fixed
telescopes remains interesting because of its very long maximum baseline of
800m. A high number of measurements at different spatial frequencies
are also expected to lead to aperture synthesis imaging. In fact even
a small number of spatial frequencies should permit to achieve
parametric images using regularization methods that takes into account
the astrophysical nature of the sources.

I would like to emphasize also that interferometers up to now provided
only broad-band visibility amplitudes. IOTA has been upgraded
to 3 telescopes and will soon provide closure phases on YSOs, an
observable which gives information on the symmetry of the obscured
structure. A centro-symmetric object has a zero closure phase, whereas
a non centro-symmetric object has a non-zero closure phase.
\inlinecite{Lac03,Malspie2000} simulated circumstellar disks irradiated
by young stars and showed that the expected closure phase with typical
100-m baselines can reach 50 degrees.  AMBER \cite{Pet02} on the VLTI
will also provide closure phases whenever used with a triplet of
auxiliary or unit telescopes. AMBER and MIDI on the VLTI will provide
for the first time spectral information on YSOs in the IR domain. With
such spectral information, relative visibility amplitudes and phases
through the spectrum and therefore through spectral lines can be
measured. This feature should bring out a radical change in the
observational constraints because it includes both morphological and
kinematics information.

\subsection{Some examples of new YSO investigations}

In this part, I present some non-exhaustive examples of new types of
observations than can be carried out with instruments providing the type of improvements listed in
the previous paragraph. 

\subsubsection{Planet-disk interaction}

\inlinecite{Wolf01} have investigated the possibility to detect the
signature of a protoplanet inside a protoplanetary disk. Some
exoplanets discovered by radial velocity techniques are massive giant
planets located close to the parent star. This location was not
predicted by theory, but since then, several models were able to
explain the presence of such planets thanks to migration mechanisms in
a protoplanetary disk \cite{Nel00}. The idea of \citeauthor{Wolf01}, was to
see if it would be possible to detect such a gap in accretion disks
that would be the signature of the presence of a giant planet. They
made the prediction for mid-IR observations (MIDI on the VLTI) and
demonstrated that the difference between a perfectly smooth disk and a
disk with a gap leads to a change in visibility amplitudes of the
order of 5\%. This number is at the limit of detection for the
instrument MIDI, but within realistic expectations for current
near-infrared instruments or  future instruments.

\subsubsection{Opening angle of microjets}
\label{sect:jets}

\inlinecite{ED97} and \inlinecite{Gar01} proposed to investigate the morphology of
jets close to the stars. Jets are an important feature of young
stellar objects, almost always present whenever there is a disk. For
the moment very little is known on the interaction between the disk,
the stars and the jets. In fact several magneto-hydrodynamical models
have been developed and optical interferometry could be the tool to
test them on very high angular resolution observations.

However, even before directly testing the physics of jets, the simple
observation of the opening angle of jets would bring significant
constrain on models. Up to now the resolution of a single dish
telecope was not sufficient to see the base of the jets. \inlinecite{Gar01}
used a toy model to simulate a hollow jet and computed the expected
visibility of a star-jet system. The expected visibility amplitude is
of the order of 50\%. Garcia et al.\ demonstrated that RU~Lupus can be
observed with sufficient signal-to-noise ratio with AMBER and medium
spectral resolution.

\subsubsection{Probing the velocity field of disks at sub-AU scale}
\label{sect:kepler}

\begin{figure}[t]
  \centering
  \includegraphics[width=0.7\hsize]{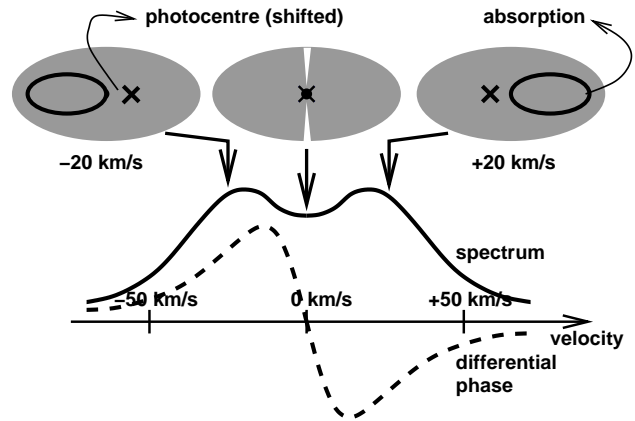}
  \caption[]{Principle to measure the Keplerian motion of matter inside a
    circumstellar disk (Lachaume et al., priv. comm.). See
    the text for details.}
  \label{fig:kepler}
\end{figure}

Spectral information is indeed the feature that could bring one the
most interesting information in YSOs. Lachaume, Malbet and Monin
(priv. comm.) proposed to observe circumstellar disks with enough
spectral resolution to be able to detect the Keplerian motion of the
matter in the disk.  \inlinecite{HKH88} have correlated several
absorption lines of the spectrum of FU Orionis in order to detect the
Doppler shift signal coming from the matter orbiting around the star.
They found a double peak line that could be interpreted as the
evidence of Keplerian motion in the disk. However they were not able
to completely rule out the presence of a spectroscopic binary in the
system. With AMBER and the spectral information, one can follow the
position of the photocenter by measuring the wavelength-dependent
phase. Figure \ref{fig:kepler} shows the principle of the measure: in
the blue part of the absorption line the matter which absorb the light
is located on one part of the disk and therefore the photocenter of
remaining light is shifted to the opposite side of the disk. For the
red part of the spectrum, the signal is shifted in the opposite
direction. In the middle of the line the disk is symmetrical and the
photocenter is aligned with the star. Therefore the phase will
increase positively in the blue part of the line goes through zero at
the systemic velocity and increase negatively in the red part. The
shape of this curve can be compared to the one expected from a pure
Keplerian motion. Any departure would be an interesting test of the
physics of the disk.  The signal will be faint, but the plan is to
correlate the signals from various spectral lines as is done for
radial velocity measurements.

\section{Conclusion}

I reviewed the science that has been performed so far
with existing interferometers. Some important information has been
obtained mainly for the most massive objects already observable. A
wider range of issues can be addressed by large interferometers since
a large fraction of YSOs will be observable with a large number of
different instrumental configurations. This astrophysical domain is
therefore undoubtedly progressing thanks to this new high angular
resolution technique and the results will be of importance not only
for star formation but also for general physical processes like the
origin of turbulence, the interraction between accretion and ejection.

However, the wavelength coverage of current and planned instruments does
not cover yet the visible part of the spectrum, the field is limited
to compact objects and imaging is still a goal. For all these reasons,
star formation remains a major scientific drivers for next generation
interferometric instruments.

\begin{acknowledgements}
  I would like to thank J.-P.~Berger, R.~Lachaume and
  R.~Millan-Gabet for very interesting discussions and inputs for this
  review that was initially prepared for the SPIE'2002 meeting. I am
  also grateful to R.~Millan-Gabet and and J.D.~Monnier who reread the
  manuscript. 
\end{acknowledgements}

\end{article}
\end{document}